\documentclass[12pt]{article}

\usepackage{amsmath}
\usepackage{stmaryrd}
\usepackage{subfigure}
\usepackage{empheq}
\usepackage{appendix,epic,eepic,amsmath,amsthm,amssymb,epsfig,cite,indentfirst,oldgerm}
\renewcommand{\theequation}{\arabic{equation}}
\newcommand{\EQ}{\begin{equation}}
\newcommand{\EN}{\end{equation}}

\newcommand{\ket}[1]{\left|#1\right\rangle}      % Ket-Zustand
\newcommand{\bear}{\begin{eqnarray}}
\newcommand{\ear}{\end{eqnarray}}
\newcommand{\bt} { \begin{tabular} }
\newcommand{\et}{ \end{tabular} }
\newcommand{\bc} { \begin{center} }
\newcommand{\ec}{ \end{center} }

\newcommand{\btb} { \begin{table} }
\newcommand{\etb}{ \end{table} }
\begin{document}

\topmargin 0pt
\oddsidemargin 5mm
\newcommand{\NP}[1]{Nucl.\ Phys.\ {\bf #1}}
\newcommand{\PL}[1]{Phys.\ Lett.\ {\bf #1}}
\newcommand{\NC}[1]{Nuovo Cimento {\bf #1}}
\newcommand{\CMP}[1]{Comm.\ Math.\ Phys.\ {\bf #1}}
\newcommand{\PR}[1]{Phys.\ Rev.\ {\bf #1}}
\newcommand{\PRL}[1]{Phys.\ Rev.\ Lett.\ {\bf #1}}
\newcommand{\MPL}[1]{Mod.\ Phys.\ Lett.\ {\bf #1}}
\newcommand{\JETP}[1]{Sov.\ Phys.\ JETP {\bf #1}}
\newcommand{\TMP}[1]{Teor.\ Mat.\ Fiz.\ {\bf #1}}

\renewcommand{\thefootnote}{\fnsymbol{footnote}}

\newpage
\setcounter{page}{0}
\begin{titlepage}
\begin{flushright}

\end{flushright}
\vspace{0.5cm}
\begin{center}
{\large Integrability of spin-$\frac{1}{2}$ fermions with charge pairing and Hubbard interaction.} \\
\vspace{1cm}
{\large M.J. Martins } \\
\vspace{0.15cm}
{\em Universidade Federal de S\~ao Carlos\\
Departamento de F\'{\i}sica \\
C.P. 676, 13565-905, S\~ao Carlos (SP), Brazil\\}
%E-mail Address: {\tt martins@df.ufscar.br}}\\
\vspace{0.35cm}
\end{center}
\vspace{0.5cm}

\begin{abstract}
In this paper we study the exact solution of a one-dimensional model 
of spin-$\frac{1}{2}$ electrons composed by a nearest-neighbor triplet
pairing term and the on-site Hubbard interaction. We argue that this
model admits a Bethe ansatz solution through a mapping to a Hubbard
chain with imaginary kinetic hopping terms. The Bethe equations are
similar to that found by Lieb and Wu \cite{LW} but with additional 
twist phases which are dependent on the ring size. We have studied the
spectrum of the model with repulsive interaction 
by exact diagonalization and through the
Bethe equations for large lattice sizes. One feature of the model
is that it is possible to define the charge gap for even and odd lattice 
sites and both converge to the same value in the infinite size limit.
We analyze the finite-size corrections
to the low-lying spin excitations and argue that they are equivalent
to that of the spin-$\frac{1}{2}$ isotropic Heisenberg model with
a boundary twist depending on the lattice parity. We present the
classical statistical mechanics model whose transfer matrix commutes
with the model Hamiltonian. To this end we have used the construction 
employed by Shastry \cite{SHA1,SHA2} for the Hubbard model. 
In our case, however,
the building block is 
a free-fermion eight-vertex model with a particular null weight.
\end{abstract}
%\vspace{.1cm} \centerline{}
\vspace{.1cm} \centerline{Keywords: Lattice fermions model, Integrability, Bethe ansatz, Yang-Baxter.}
\vspace{.1cm} \centerline{ June 2020}

\end{titlepage}

%\tableofcontents

\pagestyle{empty}

\newpage

\pagestyle{plain}
\pagenumbering{arabic}

\renewcommand{\thefootnote}{\arabic{footnote}}
\newtheorem{proposition}{Proposition}
\newtheorem{pr}{Proposition}
\newtheorem{remark}{Remark}
\newtheorem{re}{Remark}
\newtheorem{theorem}{Theorem}
\newtheorem{theo}{Theorem}

\def\ll{\left\lgroup}
\def\rr{\right\rgroup}

\newtheorem{Theorem}{Theorem}[section]
\newtheorem{Corollary}[Theorem]{Corollary}
\newtheorem{Proposition}[Theorem]{Proposition}
\newtheorem{Conjecture}[Theorem]{Conjecture}
\newtheorem{Lemma}[Theorem]{Lemma}
\newtheorem{Example}[Theorem]{Example}
\newtheorem{Note}[Theorem]{Note}
\newtheorem{Definition}[Theorem]{Definition}

\section{The Model Hamiltonian}

In general, correlations among fermions in one-dimension 
give rise to complex phase diagram with charge and 
spin ordering. One of the simplest lattice system model
that describes the effect of
such correlations is 
the Hubbard model \cite{GUZ,HUB}. This model encodes the
basics physics concerning the competition between
electron kinetic energy and the on-site Coulomb 
interaction denoted here by $U$. The
Hamiltonian of this model on a ring of size $L$ with 
a electron-hole 
symmetric interaction
is given by,
\begin{equation}
\label{HAM}
H=-\sum_{j=1}^{L} \sum_{\alpha=\uparrow,\downarrow} 
\left[c^{\dagger}_{\alpha}(j)c_{\alpha}(j+1) + c^{\dagger}_{\alpha}(j+1)c_{\alpha}(j) \right]
+U \sum_{j=1}^{L}(n_{\uparrow}(j)-\frac{1}{2})(n_{\downarrow}(j)-\frac{1}{2}) ,  
\end{equation}
where $c^{\dagger}_{\alpha}(j)$ and $c_{\alpha}(j)$ creates and annihilates 
fermions on site $j$
with spin $\alpha$ and $n_{\alpha}(j)=c^{\dagger}_{\alpha}(j)c_{\alpha}(j)$ 
is the occupation
number operator. Here we apply periodic boundary 
conditions by identifying the sites $L+1 \equiv 1$.  

In 1968 Lieb and Wu showed that the Hamiltonian (\ref{HAM}) can be diagonalized
by an extension of
the Bethe ansatz technique \cite{LW}. 
They used this solution to argue
that the Hubbard model at half-filling is an insulator for positive values of $U$ 
and undergoes a 
Mott transition at $U=0$. The literature exploring the solution by Lieb
and Wu is nowadays vast and for a collection of reprints and an extensive 
review on this
subject see for instance
\cite{MON,EKG}. In the context of this paper we mention
the progresses made by Shastry towards the understanding 
of the algebraic
structure associated to the integrability 
of the
one-dimensional Hubbard model \cite{SHA1,SHA2}. In particular, this author
discovered a two-dimensional vertex model of 
classical statistical mechanics
whose transfer matrix commutes among themselves and with 
the Hubbard Hamiltonian.

The purpose of this work is to introduce a variant 
of the Hubbard model
and to discuss its solution by the Bethe ansatz as well 
as to uncover the underlying
covering vertex model. The model is defined replacing
the hopping term of the Hubbard chain by
a nearest-neighbor charge pairing potential. 
The corresponding model Hamiltonian is,
\begin{equation}
\label{HAMC}
H_c=\sum_{j=1}^{L} \sum_{\alpha=\uparrow,\downarrow} \left[
c_{\alpha}(j)c_{\alpha}(j+1) + c^{\dagger}_{\alpha}(j+1)c^{\dagger}_{\alpha}(j) \right]
+U \sum_{j=1}^{L}(n_{\uparrow}(j)-\frac{1}{2})(n_{\downarrow}(j)-\frac{1}{2}),  
\end{equation}
where periodic boundary conditions are assumed.

We observe that the first term of Hamiltonian (\ref{HAMC})
causes charges 
to be created or annihilated
in pairs being similar to a triplet pairing in the 
p-wave theory of superconductivity. Here we are considering 
the situation
in which the pairing energy
is the same for both spin 
up and down channels. The interaction term is the same 
as that of the Hubbard model
which is taken 
symmetric under the electron-hole transformation
$c_{\alpha}(j) \leftrightarrow c^{\dagger}_{\alpha}(j)$. 

The charge pair model (\ref{HAMC}) enjoys of translation
invariance, 
the symmetry under spin flips and the invariance
under two
$Z_2$ symmetries represented by the 
unitary transformation $V_{\sigma} H_c V_{\alpha}^{\dagger}$ with 
$V_{\alpha}=e^{i\pi \sum_{j=1}^{L} n_{\alpha}(j)}$.
Besides that we have other global invariance with 
respect to specific rotations associated to the spin space.
In order to describe that we first
recall the structure of the
isomorphic $SU(2)$ algebras which can be constructed 
out of
the possible six non-vanishing on-site combinations 
of spin-$\frac{1}{2}$ fermionic
operators. The on-site generators of the standard spin
$SU(2)$ algebra is known to be given by,
\begin{equation}
\label{SPIN}
S_j^{x}=
\frac{1}{2}[c^{\dagger}_{\uparrow}(j)c_{\downarrow}(j)+
c^{\dagger}_{\downarrow}(j)c_{\uparrow}(j)],~~~ 
S_j^{y}=
\frac{i}{2}[c^{\dagger}_{\downarrow}(j)c_{\uparrow}(j)-
c^{\dagger}_{\uparrow}(j)c_{\downarrow}(j)],~~~ 
S_j^{z}=
\frac{1}{2}[n_{\uparrow}(j)-n_{\downarrow}(j)].
\end{equation}

Yet another basis can be obtained by applying for instance
the electron-hole transformation on the spin down 
fermionic operators
defined by (\ref{SPIN}). This gives rises to 
the so-called pseudo-spin or charge
$SU(2)$ algebra whose on-site generators are,
\begin{equation}
\label{SUDO}
R_j^{x}=
\frac{1}{2}[c^{\dagger}_{\uparrow}(j)c^{\dagger}_{\downarrow}(j)+
c_{\downarrow}(j)c_{\uparrow}(j)],~~~ 
R_j^{y}=
\frac{i}{2}[c_{\downarrow}(j)c_{\uparrow}(j)-
c^{\dagger}_{\uparrow}(j)c^{\dagger}_{\downarrow}(j)],~~~ 
R_j^{z}=
\frac{1}{2}[n_{\uparrow}(j)+n_{\downarrow}(j)-1].
\end{equation}

For arbitrary values of $L$ the Hubbard model (\ref{HAM}) 
is invariant by the 
full non-Abelian spin $SU(2)$ symmetry (\ref{SPIN}).
However, this is not the case of the 
Hamiltonian (\ref{HAMC}) which is invariant only
when this symmetry is broken down to rotations 
around the $y$ axis. The same observation 
applies to the pseudo-spin
$SU(2)$ algebra (\ref{SUDO}) since the  
charge pair model is invariant 
by such symmetry when
it is restricted to rotations around the $x$ axis. 
More precisely,
for arbitrary values of $L$ we have the following 
conservation laws\footnote{For $L$ even we will see that such rotations 
around specific axes
are enlarged to $SU(2)$ symmetries.}
\begin{equation}
\label{INV}
[H_c,\sum_{j=1}^{L} S_j^y]=[H_c, \sum_{j=1}^{L}R_j^x]=0.
\end{equation}

In next section we shall explore the commutations (\ref{INV}) 
in order to 
determine the eigenspectrum of the model
by the coordinate nested Bethe ansatz method. 
These charges 
can be made
equivalent to the conservation of particle numbers 
by means of a redefinition 
of the original electrons operators. In this new fermionic
basis, the Hamiltonian (\ref{HAMC}) is 
mapped to the form of the
Hubbard model but with pure imaginary 
and asymmetric hopping terms. We find that 
the corresponding Bethe
equations for $L$ even are distinct from 
that with $L$ odd 
through suitable boundary twists.  In section 3 we investigate
the properties of the spectrum of the model for repulsive 
interactions. We argue that for both $L$ even and odd we
can define a lattice charge gap with respect to the 
ground state which
in the thermodynamic limit converges to the value computed
by Lieb and Wu for the Hubbard model at half-filling \cite{LW}.
We have used the Bethe equations to study the finite-size
corrections associated to the excitations due
to the spin degrees of freedom. We find that they are equivalent
to that of the isotropic spin-$\frac{1}{2}$ Heisenberg model with periodic boundary 
for $L$ even and with a twisted toroidal boundary when $L$ is odd.
In section 4 we describe 
the lattice vertex model whose transfer matrix commutes with
the Hamiltonian (\ref{HAMC}). This is done by using a
construction
due to Shastry devised to couple two symmetric six-vertex models 
satisfying
the free-fermion condition \cite{SHA1,SHA2}. However, in our case
the building block has the form of an eight-vertex model in which
one of the weights is zero.
The fact
that Shastry's formulation also works for such
special $Z_2$ invariant vertex model seems to have been unnoticed
in the literature. Our concluding remarks are given in section 5 
and in Appendix A we present the technical details on the underlying
Yang-Baxter algebra.

\section{The Energy Hamiltonian Spectrum}

The space of states of spin-$\frac{1}{2}$ fermions associated with every 
lattice is four-dimensional
and they can be represented as,
\begin{equation}
\label{basis}
\ket{0},~~~c^{\dagger}_{\uparrow}(j)\ket{0},~~~c^{\dagger}_{\downarrow}(j)\ket{0},~~~
c^{\dagger}_{\uparrow}(j)c^{\dagger}_{\downarrow}(j)\ket{0},
\end{equation}
where $\ket{0}$ denotes the vacuum state defined by the condition
$c_{\alpha}(j)\ket{0}=0$.

In the above canonical basis the local conserved charges 
$S_j^{y}$ and $L_j^{x}$ are 
viewed as anti-diagonal matrices. However, they can be both 
diagonalized 
by on-site
unitary transformation with the following similarity matrix,
\begin{equation}
\label{UNI}
V_j=\frac{1}{\sqrt{2}}\left(
\begin{array}{cccc}
1& 0& 0& 1 \\
0& -i& 1& 0 \\
0& -1& i& 0 \\
-1& 0& 0& 1 \\
\end{array}
\right)_j.
\end{equation}

We can use this transformation to define new 
on-site fermionic 
operators,
\begin{equation}
d_{\alpha}(j) =V_j c_{\alpha}(j) V^{\dagger}_j,~~~
d^{\dagger}_{\alpha}(j) =V_j c^{\dagger}_{\alpha}(j) V^{\dagger}_{j},
\end{equation}
and their explicit expressions in terms of the  
electrons operators are,
\begin{eqnarray}
\label{TRAN}
&& d_{\downarrow}(j) = \frac{i}{2} c_{\uparrow}(j)+\frac{1}{2} c^{\dagger}_{\uparrow}(j) 
-\frac{1}{2} c_{\downarrow}(j) +\frac{i}{2} c^{\dagger}_{\downarrow}(j),~~~ 
d^{\dagger}_{\downarrow}(j) = -\frac{i}{2} c^{\dagger}_{\uparrow}(j)+\frac{1}{2} c_{\uparrow}(j) 
-\frac{1}{2} c^{\dagger}_{\downarrow}(j) -\frac{i}{2} c_{\downarrow}(j), 
\nonumber \\ \nonumber \\
&& d_{\uparrow}(j) = \frac{1}{2} c_{\uparrow}(j)+\frac{i}{2} c^{\dagger}_{\uparrow}(j) 
-\frac{i}{2} c_{\downarrow}(j) +\frac{1}{2} c^{\dagger}_{\downarrow}(j),~~~ 
d^{\dagger}_{\uparrow}(j) = \frac{1}{2} c^{\dagger}_{\uparrow}(j)-\frac{i}{2} c_{\uparrow}(j) 
+\frac{i}{2} c^{\dagger}_{\downarrow}(j) +\frac{1}{2} c_{\downarrow}(j). \nonumber  \\
\end{eqnarray}

By transforming back the above relations we can represent the conserved charges 
in terms of
the new fields 
$d_{\alpha}(j)$  and 
$d^{\dagger}_{\alpha}(j)$. The expression of the spin algebra charge component is,
\begin{equation}
\label{CONS1}
\sum_{j=1}^{L}S^y_j=\frac{1}{2}\sum_{j=1}^{L}\left[d^{\dagger}_{\uparrow}(j) d_{\uparrow}(j)-
d^{\dagger}_{\downarrow}(j) d_{\downarrow}(j)\right],
\end{equation}
while the one associated to the pseudo-spin algebra is,
\begin{equation}
\label{CONS2}
\sum_{j=1}^{L}R^x_j=\frac{1}{2}\sum_{j=1}^{L}\left[d^{\dagger}(j)_{\uparrow}(j) d_{\uparrow}(j)+
d^{\dagger}_{\downarrow}(j) d_{\downarrow}(j)-1\right] .
\end{equation}

By the same token 
the Hamiltonian (\ref{HAMC}) of the charge pair model can be expressed as follows,
\begin{eqnarray}
\label{HAMCTRA}
\tilde{H}_c&=&\sum_{j=1}^{L}\left[ 
e^{i\pi/2}d^{\dagger}_{\uparrow}(j)d_{\uparrow}(j+1) +e^{-i\pi/2} d^{\dagger}_{\uparrow}(j+1)d_{\uparrow}(j) 
+e^{-i\pi/2}d^{\dagger}_{\downarrow}(j)d_{\downarrow}(j+1) +e^{i\pi/2} d^{\dagger}_{\downarrow}(j+1)d_{\downarrow}(j) \right]
\nonumber \\
&+&U \sum_{j=1}^{L}(d^{\dagger}_{\uparrow}(j)d_{\uparrow}(j)-\frac{1}{2})  
(d^{\dagger}_{\downarrow}(j)d_{\downarrow}(j)-\frac{1}{2}) , 
\end{eqnarray}
which has the typical form of the Hubbard Hamiltonian 
however with imaginary 
and asymmetric hopping terms.

The conserved charges (\ref{CONS1},\ref{CONS2}) imply that
the Hilbert space of the Hamiltonian (\ref{HAMCTRA}) can be separated into 
block disjoint sectors
labeled by the total number 
$N_{\alpha}=\sum_{j=1}^{L}d^{\dagger}_{\alpha}(j)d_{\alpha}(j)$
of fermions of spin $\alpha$. This means that
eigenvalue problem can be formally written as, 
\begin{equation}
\tilde{H}_c \ket{N_{\uparrow},N_{\downarrow}}= E(N_{\uparrow},N_{\downarrow},U) \ket{N_{\uparrow},N_{\downarrow}}.
\end{equation}

The range of the quantum numbers can be constrained observing
that Hamiltonian (\ref{HAMCTRA}) is invariant under the particle-hole
symmetry
$d_{\alpha}(j) \leftrightarrow d^{\dagger}_{\alpha}(j)$. In fact, taking into account 
this invariance
we obtain the spectral identity,
\begin{equation}
\label{rel}
E(N_{\uparrow},N_{\downarrow},U) =
E(L-N_{\uparrow},L-N_{\downarrow},U).
\end{equation}

Here we remark that such spectral relation is valid 
for arbitrary $L$
in the case of the transformed charge pair 
model Hamiltonian (\ref{HAMCTRA}). Therefore, unlike 
the Hubbard model
no restriction 
to bipartite lattices is necessary in order to relate the energies of
different sectors with 
the same coupling $U$ \cite{LW}. 
The spectral identity (\ref{rel})
together with spin flip invariance tell us that we may restrict our 
considerations to states, 
\begin{equation}
\label{cons}
N_{\uparrow} +N_{\downarrow} \leq L~~~\mathrm{and}~~~N_{\uparrow} \geq N_{\downarrow},
\end{equation}
for even and odd values of $L$. We emphasize that  in the case of the 
Hubbard model (\ref{HAM}) the spectral 
relation (\ref{rel}) and
constraint (\ref{cons}) only works when L is even \cite{LW}.

We now can determine the eigenspectrum of the 
Hamiltonian (\ref{HAMCTRA}) 
by adapting the nested Bethe ansatz
approach employed Lieb and Wu \cite{LW} in the presence of
hopping phases. 
In a given sector with total number of 
fermions $N=N_{\uparrow}+N_{\downarrow}$ 
the wave function may be represented as
linear combination of $N$-particle states,
\begin{equation}
\label{wave}
\ket{N_{\uparrow},N_{\downarrow}}= \sum_{\stackrel{x_1, x_2,\dots,x_N=1}{\alpha_1,\alpha_2,\dots,\alpha_N=\uparrow,\downarrow}}^{L} 
\psi_{\alpha_1,\alpha_2,\dots,\alpha_N}(x_1,x_2,\dots,x_N) \prod_{j=1}^{N}
e^{i\phi_{\alpha_j}(1-x_j)} 
d^{\dagger}_{\alpha_j}(x_j)  
\ket{\tilde{0}}, 
\end{equation}
where the reference state $\ket{\tilde{0}}$ is taken 
such that $d_{\alpha}(j) \ket{\tilde{0}}=0$ for any 
site $j$ and spin $\alpha$. The exponentials terms in (\ref{wave})
are able to pull the hopping bulk phases up to the boundary terms.
In our case this happens when we choose the twists to be fixed as
$\phi_{\uparrow}=\frac{\pi}{2}$
and $\phi_{\downarrow}=-\frac{\pi}{2}$.

In the Bethe ansatz approach one assumes that the $N$-particle amplitudes have 
a plane wave form \cite{LW},
\begin{equation}
\psi_{\alpha_1,\alpha_2,\dots,\alpha_N}(x_1,x_2,\dots,x_N) = \sum_{P} A(Q|P) 
\exp[ik_{P_1}x_{Q_1}
+ik_{P_2}x_{Q_2}+\dots+
ik_{P_N}x_{Q_N}],
\end{equation}
where it is assumed the ordering $x_{Q_1} \leq x_{Q_2} \leq \dots \leq x_{Q_N}$.
The partition $Q=\{Q_1,Q_2,\dots,Q_N\}$ denotes the 
$N!$ permutations of fermions with positions $x_{Q_1},x_{Q_2},\dots,x_{Q_N}$ 
and spins $\alpha_1,\alpha_2,\dots,\alpha_N$  while
$P=\{P_1,P_2,\dots,P_N\}$ refers to similar permutations on the 
fermions momenta
$k_{P_1},k_{P_2},\dots,k_{P_N}$. The coefficients associated to these
permutations are denoted by $A(Q|P)$.

In this formulation the hopping phases are all removed excepted those associated 
to fermions
hopping among the boundary sites $j=1$ and $j=L$. At this point the situation becomes 
equivalent 
to that 
of generalized diagonal boundary conditions discussed for the 
Hubbard model in \cite{SASU,MAFY}. This fact is
taken into account by requiring that
$A(Q,P)$ satisfy the condition,
\begin{equation}
\exp(i k_{P_N} L) A(Q|P)= \left[\exp(i\frac{\pi L}{2}) \delta_{Q_N,N} +
\exp(-i\frac{\pi L }{2}) (1-\delta_{Q_N,N}) \right] A(\bar{Q}|\bar{P}),
\end{equation}
where $\bar{Q}=\{Q_N,Q_1,\dots,Q_{N-1}\}$ and 
$\bar{P}=\{P_N,P_1,\dots,P_{N-1}\}$ are cyclic permutations 
of the partitions $Q$ and $P$, respectively.

From now on the procedure is analog to that already exposed by
Lieb and Wu \cite{LW} and we shall present only the main results.
The spectrum of the Hamiltonian (\ref{HAMCTRA}) is parametrized 
in terms
of a set of variables $\{k_j,\mu_j\}$ which fulfill 
the following
nested Bethe equations,
\begin{eqnarray}
\label{Bethe1}
&& e^{ik_jL} = e^{\frac{i\pi L}{2}}
\prod_{l=1}^{N_{\downarrow}}  \frac{\sin(k_j) - {\mu}_{l} + \frac{iU}{4}}{\sin(k_j) - {\mu}_{l} -\frac{iU}{4}},~~~j=1,2,\dots,N_{\uparrow}+N_{\downarrow},
\\
\label{Bethe2}
&& \prod_{j=1}^{N_{\uparrow}+N_{\downarrow}}  \frac{\sin(k_j) - {\mu}_{l} + \frac{iU}{4}}{\sin(k_j) - {\mu}_{l} - \frac{iU}{4}}  =
e^{i\pi L }\prod_{\stackrel{k=1}{k \neq j}}^{N_{\downarrow}} \frac{{\mu}_{l}-{\mu}_{k}-\frac{iU}{2}}{{\mu}_{l}-{\mu}_{k} 
+\frac{iU}{2}},~~~l=1, \ldots, N_{\downarrow},
\end{eqnarray}
while the eigenvalue of the transformed Hamiltonian (\ref{HAMCTRA}) associated with the state 
specified by the rapidities $\{k_j,\mu_j\}$ is given by,
\begin{equation}
\label{EIN}
E(N_{\uparrow},N_{\downarrow},U) = -2\sum_{j=1}^{N_{\uparrow}+N_{\downarrow}} \cos(k_j)+\frac{U}{2}(\frac{L}{2}-N_{\uparrow}-N_{\downarrow}).
\end{equation}

We would like to close this section with the following comments.  
We first observe that
the fermionic chain 
for $L=2$ is somehow
special since the charge
pairing terms are canceled and the Hamiltonian 
(\ref{HAMC}) becomes
a diagonal operator. From the Bethe solution point of view
this peculiarity is associated with the presence of the  
minus sign factor in the first level Bethe equation (\ref{Bethe1}).  
We have checked this fact by solving 
the two sites Bethe equations (\ref{Bethe1},\ref{Bethe2}) for roots configurations
satisfying the restriction (\ref{cons}). These solutions
indeed reproduce the expected Hamiltonian energies and our findings
have been summarized 
in Table \ref{TAB2}.
\begin{table}[ht]
\begin{center}
\begin{tabular}{|c|c|c|} \hline
$(N_{\uparrow},N_{\downarrow})$ & $E(N_{\uparrow},N_{\downarrow},U)$ & $\mathrm{Bethe~roots}$ \\ \hline 
$(0,0)$ & $\frac{U}{2}$ & $\mathrm{empty~set}$\\ \hline 
$(1,0)$ & $0$  
& $k_1=\pm\frac{\pi}{2}$ \\ \hline 
$(2,0)$ & -$\frac{U}{2}$ & $k_1=\frac{\pi}{2},~k_2=-\frac{\pi}{2} $\\ \hline 
$(1,1)$ & $\frac{U}{2}$ &
$e^{ik_1}=\frac{-U-\sqrt{U^2-16}}{4},~
e^{ik_2}=\frac{-U+\sqrt{U^2-16}}{4},~\mu_1=0$
\\ \hline
$(1,1)$ & -$\frac{U}{2}$ & 
$k_1=0,~k_2=\pi,~\mu_1=0$ \\ \hline
\end{tabular}
\caption{The spectrum of Hamiltonians (\ref{HAMC},\ref{HAMCTRA}) for $L=2$  where 
the sectors $(N_{\uparrow},N_{\downarrow})$ satisfy (\ref{cons}). 
The
eigenvalues are obtained substituting the Bethe roots into 
the relation (\ref{EIN}).}
\label{TAB2}
\end{center}
\end{table}

We next note that for arbitrary $L$ the 
Bethe equations (\ref{Bethe1},\ref{Bethe2}) are similar  
to that of the Hubbard model and the main difference 
are the presence 
of certain phase factors depending on the lattice parity. 
In particular,
when $L$ is multiple of four such phase factors are unity resulting in 
the same Bethe equations of the Hubbard model. We conclude 
that for $L=4,8,12,\dots$ 
the spectrum of the Hubbard Hamiltonian (\ref{HAM}) and the charge 
pair model (\ref{HAMC}) should be exactly the same. This fact has been verified
for $L=4,8$ by comparing all the energy levels of both Hamiltonians
using exact diagonalization.
However, the structure of the wave-function on the electron basis is expected
to be rather different because of the canonical transformation (\ref{TRAN}).
We can see that considering examples of simple states whose energy per  site
is independent of the size $L$. From the Hubbard model perspective we already 
know that there exist
two such eigenvalues associated to the trivial 
ferromagnetic and anti-ferromagnetic states. We find out that these energies
also belong to the spectrum of 
the charge pair model Hamiltonian (\ref{HAMC}) but with distinct wave-function
structure. The form of the wave-function on the canonical basis can be uncovered 
with the help of the transformation (\ref{TRAN}). The final results for such states
have been summarized on Table \ref{TAB1}. 
\begin{table}[ht]
\begin{center}
\begin{tabular}{|c|c|c|} \hline
$\mathrm{Eigenvalue}$ & $\mathrm{Hubbard~Eigenvector}$ & $\mathrm{Charge~Pair~Eigenvector}$ \\ \hline 
$\frac{LU}{4}$ & $\ket{0},~\displaystyle \prod_{j=1}^{L} c^{\dagger}_{\uparrow}(j) c^{\dagger}_{\downarrow}(j) \ket{0}$ & 
$ \displaystyle \prod_{j=1}^{L} \left(1 \pm c^{\dagger}_{\uparrow}(j) c^{\dagger}_{\downarrow}(j)\right) \ket{0}$  
\\ \hline 
-$\frac{LU}{4}$ & $\displaystyle \prod_{j=1}^{L} c^{\dagger}_{\uparrow}(j)\ket{0},~ \displaystyle \prod_{j=1}^{L}c^{\dagger}_{\downarrow}(j) \ket{0}$ & 
$\displaystyle \prod_{j=1}^{L} \left(c^{\dagger}_{\uparrow}(j) \pm i c^{\dagger}_{\downarrow}(j) \right) \ket{0}$  
\\ \hline 
\end{tabular}
\caption{Example of eigenstates 
the Hubbard model 
(\ref{HAM}) and the charge pair 
model (\ref{HAMC}) with common eigenvalue for arbitrary $L$.} 
\label{TAB1}
\end{center}
\end{table}

Finally, we remark that for a bipartite lattice the rotation 
invariance (\ref{INV}) 
of the charge pair
Hamiltonian around specific axes is enlarged to 
the invariance under two $SU(2)$ symmetries. This is similar
to the case of the Hubbard model (\ref{HAM}) which for $L$ even 
has besides the spin $SU(2)$ symmetry (\ref{SPIN}) another
distinct $SU(2)$ invariance 
named the $\eta$-pairing symmetry \cite{YY,PER}.

In fact, for an even number of lattice sites the invariance 
of the Hamiltonian (\ref{HAMC}) under the
rotation around the y-axis of the spin algebra (\ref{SPIN}) extends
to a full ``spin" $SU(2)$ symmetry, namely
\begin{equation}
[H_c,\sum_{j=1}^{L} \tilde{S}_j^x]=[H_c, \sum_{j=1}^{L}S_j^y]=
[H_c,\sum_{j=1}^{L} \tilde{S}_j^z]=0
\end{equation}
where now the extra on-site generators 
$\tilde{S}_j^{x}$ and ${\tilde{S}}_j^{z}$ alternate
among the lattice sites,
\begin{equation}
\tilde{S}_j^{x}=
\frac{1}{2}(-1)^{j}[c^{\dagger}_{\uparrow}(j)c_{\downarrow}(j)+
c^{\dagger}_{\downarrow}(j)c_{\uparrow}(j)],~~~ 
\tilde{S}_j^{z}=
\frac{1}{2}(-1)^{j}[n_{\uparrow}(j)-n_{\downarrow}(j)].
\end{equation}

The same happens to the rotation around the x-axis of 
the charge algebra (\ref{SUDO}).
For $L$ even it is enlarged to the following ``charge" 
$SU(2)$ symmetry,
\begin{equation}
[H_c,\sum_{j=1}^{L} {R}_j^x]=[H_c, \sum_{j=1}^{L} \tilde{R}_j^y]=
[H_c,\sum_{j=1}^{L} \tilde{R}_j^z]=0
\end{equation}
where the expression for the additional staggered 
on-site generators
$\tilde{R}_j^{y}$ and ${\tilde{R}}_j^{z}$ are given by,
\begin{equation}
\tilde{R}_j^{y}=
\frac{i}{2}(-1)^{j}[c_{\downarrow}(j)c_{\uparrow}(j)-
c^{\dagger}_{\uparrow}(j)c^{\dagger}_{\downarrow}(j)],~~~ 
\tilde{R}_j^{z}=
\frac{1}{2}(-1)^{j}[n_{\uparrow}(j)+n_{\downarrow}(j)-1].
\end{equation}

\section{The Spectrum Properties for $U>0$}

As far as the energy spectrum is concerned the
difference among the charge pair model (\ref{HAMC}) and the 
Hubbard chain (\ref{HAM}) is the presence of 
size dependent
twists in the Bethe equations. However,
these fluxes are not expected to affect the value
of ground state energy per site in the
thermodynamic limit. The value
should be
same as that of the Hubbard model 
in the half-filled case
determined long ago by Lieb and Wu \cite{LW}. Denoting this energy by
$e_{\infty}$ we have,
\begin{equation}
e_{\infty}= -4 \int_{0}^{\infty} \frac{J_0(x) J_1(x)}{x\left[\exp(Ux/2)+1 \right]}dx -\frac{U}{4}
\end{equation}
where $J_0(x)$ and $J_1(x)$ are Bessel functions.

The other basic feature of the half-filled 
Hubbard model is the presence of energy gap 
in the charge excitation sector. For $L$ even 
this mass gap was defined by Lieb and Wu \cite{LW}
as the energy $\Delta(L)$ of a particle or
a hole excitation with respect to 
the half-filled state. In the thermodynamic
limit its value was computed to be \cite{LW},
\begin{equation}
\label{gap}
\Delta({\infty})=4\int_{0}^{\infty} \frac{J_1(x)}{x\left[\exp(Ux/2)+1 \right]}dx +\frac{U}{2}-2
\end{equation}

We shall argue that for the charge
pair model (\ref{HAMC}) it is possible
to define the energy gap for
both even and odd lattice sites. It turns out 
that the phase factors for $L$ odd
compensate the effects of frustration due to the 
lattice parity
and the energy gap of either a hole or a particle 
excitation over the double degenerated
ground state is the same. We shall present numerical
evidences that the value of the gap 
for even and odd sites 
converges in the thermodynamic 
limit to the result (\ref{gap}).

The other known feature of the
Hubbard model at half-filling is
that the spin excitations are gapless
in the repulsive
regime. The phases twists for 
the charge pair model will not change
this behavior but the conformal data
will be dependent on the parity of the
lattice size. 
In what follows
we will also study the finite-size effects
for some of the gapless states
of the charge pair Hamiltonian (\ref{HAMC}).

\subsection{Finite-size effects for L even}

From the Bethe solution we concluded that the 
energy spectrum of 
the charge pair model and
the Hubbard model coincides for $L/2$ even. However, when $L/2$ is odd
the energy spectrum of these two models are not the same due to 
the presence of a minus
sign in the first Bethe equation (\ref{Bethe1}). In what follows we shall therefore
restrict our analysis of the spectrum 
for lattice sites not multiple
of four. 

\begin{figure}[ht]
\begin{center}    
\includegraphics[width=16cm]{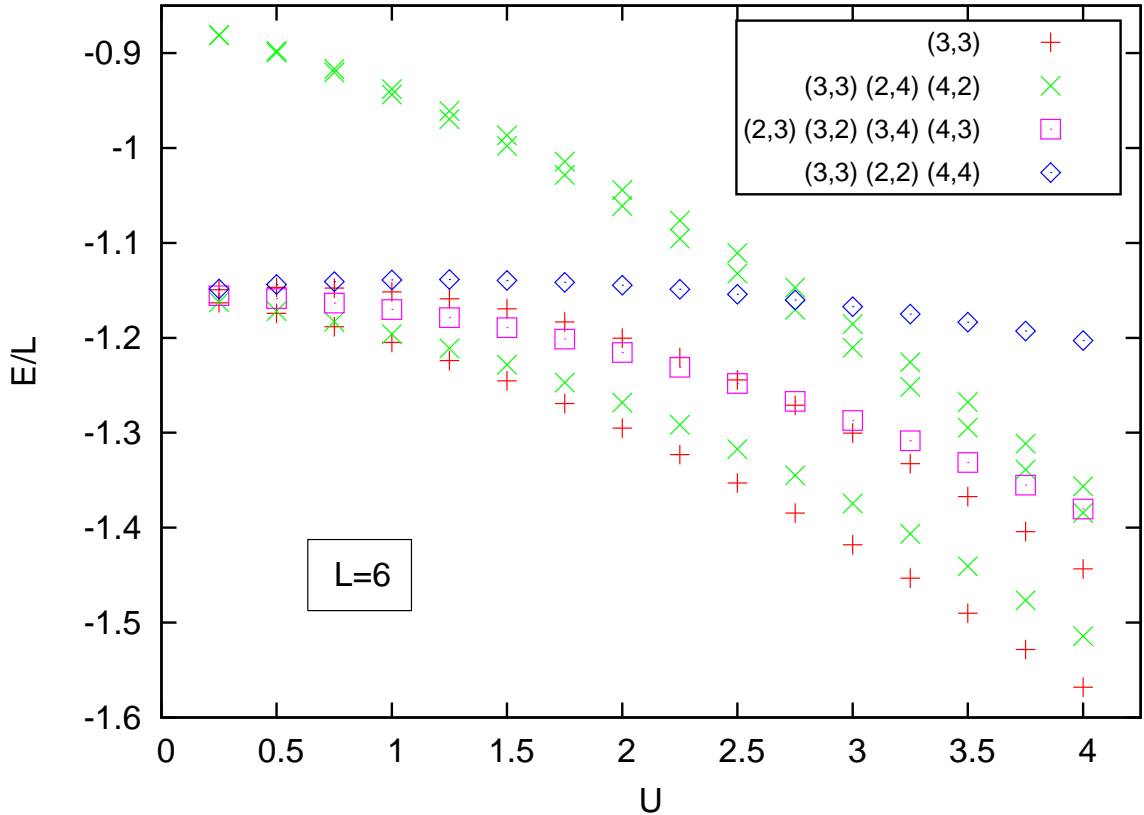}
\end{center}
\caption{The low-lying energies of Hamiltonian (\ref{HAMC}) for $L=6$ as 
function of the interaction parameter $U$. 
The energies are labeled using the quantum numbers  
$(N_{\uparrow},N_{\downarrow})$ of the Bethe 
equations (\ref{Bethe1},\ref{Bethe2}).}
\label{figL6}
\end{figure}

From the exact diagonalization of the 
Hamiltonian (\ref{HAMC}) we conclude 
that the ground state is a singlet and lies 
in the sector $(\frac{L}{2},\frac{L}{2})$.
The situation is similar to that of the Hubbard model
at half-filling.
In  Figure \ref{figL6} we exhibit the low-lying 
energies per site 
for $L=6$ in which the states are label 
using the quantum numbers associated with
the Bethe ansatz solution of the
transformed Hamiltonian (\ref{HAMCTRA}).

We find out that the ground states as well as many 
of the low-lying excitations can be described
by real roots of the Bethe ansatz 
equations (\ref{Bethe1},\ref{Bethe2}). Based on 
this observation 
we can take the logarithm of the Bethe equations
to obtain,
\begin{eqnarray}
\label{Bethereal}
&& L k_j= 2\pi Q_j^{(1)} -2\sum_{l=1}^{N_{\downarrow}} \arctan\left[\frac{\sin(k_j)-\mu_l}{U/4}  \right],~~~
j=1,\dots,N_{\uparrow}+N_{\downarrow}, \nonumber \\
&& 2\sum_{l=1}^{N_{\uparrow}+N_{\downarrow}} \arctan\left[\frac{\mu_j-\sin(k_l)}{U/4}  \right]=
2\pi Q_j^{(2)} +2\sum_{\stackrel{l=1}{l \neq j}}^{N_{\downarrow}} \arctan\left[\frac{\mu_j-\mu_l}{U/2}  \right],~~~
j=1,\dots,N_{\downarrow}
\end{eqnarray}
where the numbers 
$Q_j^{(1,2)}$ define the many possible branches of the logarithm.

In Table (\ref{TAB3}) we give the numbers 
$Q_j^{(1,2)}$ for the ground state and lowest states 
associated to 
the charge and spin sectors.
We remark that such sequence of numbers are not 
the same as that of the corresponding
states of the Hubbard model\footnote{For the Hubbard models such Bethe numbers 
depend on whether $L/2$ is even
or odd see for instance \cite{TAKA}.}
because of the extra sign on the Bethe equation (\ref{Bethe1}).
\begin{table}[ht]
\begin{center}
\begin{tabular}{|c|c|c|c|} \hline
$(N_{\uparrow},N_{\downarrow})$ & $Q_j^{(1)}$ & $Q_j^{(2)}$ & $\mathrm{state}$ \\ \hline 
$(\frac{L}{2},\frac{L}{2})$ & $\frac{L}{2}-j+1$ & -$\frac{(L-2)}{4}+j-1$ & $\mathrm{ground}~\mathrm{state}$ \\ \hline 
$(\frac{L}{2},\frac{L}{2}-1)$ & $\frac{L-1}{2}-j+1$  & -$\frac{(L-2)}{4}+j-1$ & $\mathrm{charge}~\mathrm{excitation}$ \\ \hline 
$(\frac{L}{2}+1,\frac{L}{2}-1)$ & $\frac{L-1}{2}-j+1$ & -$\frac{(L-4)}{4}+j-1$ & $\mathrm{spin}~\mathrm{excitation}$ \\ \hline 
\end{tabular}
\caption{The Bethe numbers $Q_j^{(1,2)}$ for the ground state and the lowest 
charge and spin excitations.}
\label{TAB3}
\end{center}
\end{table}

We now start to report on the numerical analysis about the 
eigenstates described in Table (\ref{TAB3}). In what follows we shall denote by
$E_j\left(N_{\uparrow},N_{\downarrow},U\right)$ 
the $j$-th energy level
in a given sector $(N_{\uparrow},N_{\downarrow})$.
The energy 
gap of one 
hole excitation 
over the singlet ground state can then be defined as,
\begin{equation}
\label{gapeven}
\Delta_{ev}(L)= E_0\left(\frac{L}{2},\frac{L}{2}-1,U\right)-E_0\left(\frac{L}{2},\frac{L}{2},U\right)
\end{equation}

In order to verify the behavior of the gap in the thermodynamic 
limit we have solved 
the Bethe equations (\ref{Bethereal}) for lattice sizes 
up to $ L = 1038$. In this solution
we have used the corresponding configurations of the numbers
$Q_j^{(1,2)}$ of Table (\ref{TAB3}). The numerical data 
for mass gap 
is presented in Table (\ref{TAB4}) together with the extrapolation 
for large lattice sizes. The extrapolated data 
is in accordance with the Lieb and Wu's result (\ref{gap}). 
\begin{table}[ht]
\begin{center}
\begin{tabular}{|c|c|c|c|} \hline
$L$ & $U=2$ & $U=3$ & $U=4$ \\ \hline 
%14   &   0.2881131678   &     0.5346153484  &   0.8411131052  \\ \hline
62   &   0.1397049178   &     0.3583388520  &   0.6783598211  \\ \hline
142  &   0.1081504685   &     0.3327705853  &   0.6577213650  \\ \hline
222  &   0.0995719373   &     0.3263342354  &   0.6523806267  \\ \hline
302  &   0.0957870201   &     0.3234174755  &   0.6499337362  \\ \hline
382  &   0.0936811316   &     0.3217548135  &   0.6485307420  \\ \hline
462  &   0.0923434712   &     0.3206809062  &   0.6476212362  \\ \hline
638  &   0.0906289820   &     0.3192820048  &   0.6464324386  \\ \hline
1038 &   0.0889500759   &     0.3178852239  &   0.6452407134  \\ \hline
$ \mathrm{Extrap.}$ & 0.08645(1) & 0.31566(1) & 0.64335(2)  \\ \hline           
$ \Delta(\infty) $ & 0.0863890951 & 0.3156965889  & 0.6433635110  \\ \hline           
\end{tabular}
\caption{The finite-size sequence $\Delta_{ev}(L)$ of the charge gap (\ref{gapeven})
for $U=2,3,4$ and the respective extrapolated value. The exact value 
is obtained from expression (\ref{gap}). }
\label{TAB4}
\end{center}
\end{table}

We now turn to the analysis of the 
finite-size corrections
to the spin degrees
of freedom. As remarked these excitations 
should be gapless since the
spectrum of the charge pair model (\ref{HAMC}) 
and the Hubbard model (\ref{HAM}) is the same
for lattice sizes multiple of four. For the Hubbard model
such spin excitations is known to show the same 
critical behavior of the isotropic spin-$\frac{1}{2}$
Heisenberg model with periodic boundary conditions \cite{WAY}.
It is therefore expected similar critical behavior for 
charge pair model (\ref{HAMC}) in the case of even number 
of lattice sizes. In particular, the
finite-size dependence of the ground state energy should
be governed by a conformal theory with central charge $c=1$.
More precisely, 
following the results obtained
by Woynarovich and Eckle \cite{WAY} one expects,
\begin{equation}
E_0\left(\frac{L}{2},\frac{L}{2},U\right)-e_{\infty}L= -\frac{\pi \xi}{6L}\Big(1+\mathcal{O}(1/\ln[LI_0(2\pi/U)]^3\Big)
\end{equation}
where $\xi =2I_1(2\pi/U)/I_0(2\pi/U)$ is the sound 
velocity of the 
spin excitation. The functions $I_0(x)$ and
$I_1(x)$ are modified Bessel functions.

We have checked the above result by numerically computing 
the estimators,
\begin{equation}
\label{central}
C(L)=\frac{6L}{\pi \xi} \left[-E_0\left(\frac{L}{2},\frac{L}{2},U\right)+e_{\infty}L\right] 
\end{equation}
for lattice sizes up to $L =1038$.
In Table (\ref{TAB5}) we have presented these estimates and we observe the rapid converge to
the expected value $c=1$. 
\begin{table}[ht]
\begin{center}
\begin{tabular}{|c|c|c|c|} \hline
$L$ & $U=2$ & $U=3$ & $U=4$ \\ \hline 
   62   &  0.6157199846    &     0.9989338450  &  1.0018587610 \\ \hline
   142  &  0.9751616589    &     1.0009382382  &  1.0010923478 \\ \hline
   222  &  0.9990148608    &     1.0007557142  &  1.0008723534 \\ \hline
   302  &  1.0004043456    &     1.0006013671  &  1.0007591646 \\ \hline
   382  &  1.0004462355    &     1.0005586587  &  1.0006872859 \\ \hline
   462  &  1.0004211966    &     1.0004958734  &  1.0006363695 \\ \hline
   638  &  1.0003772301    &     1.0004958734  &  1.0005618437 \\ \hline
   1038  & 1.0003209451    &     1.0004190615  &  1.0004712889 \\ \hline
$ \mathrm{Extrap.}$ & 1.0002(1) & 1.0003(1)    & 1.0003(2)  \\ \hline           
\end{tabular}
\caption{The finite-size sequence $C(L)$ of the central 
charge (\ref{central})  for 
$U=2,3,4$ together with the respective extrapolation for large systems.  
The predicted value for the central charge is $c=1$.}  
\label{TAB5}
\end{center}
\end{table}

From the above information we conclude that for $L$ even 
the leading behavior 
of the finite-size corrections of charge pair model should be
same of that discussed  Woynarovich and Eckle \cite{WAY} for the 
Hubbard model. As far as finite-size effects are concerned
the difference among these two models for $L/2$ odd 
appears to be associated
with the subleading corrections. The amplitudes of subleading terms are 
probably affected 
by presence of distinct 
phase factors in the first level Bethe equation. 

\subsection{Finite-size effects for L odd}

\begin{figure}[ht]
\begin{center}    
\includegraphics[width=16cm]{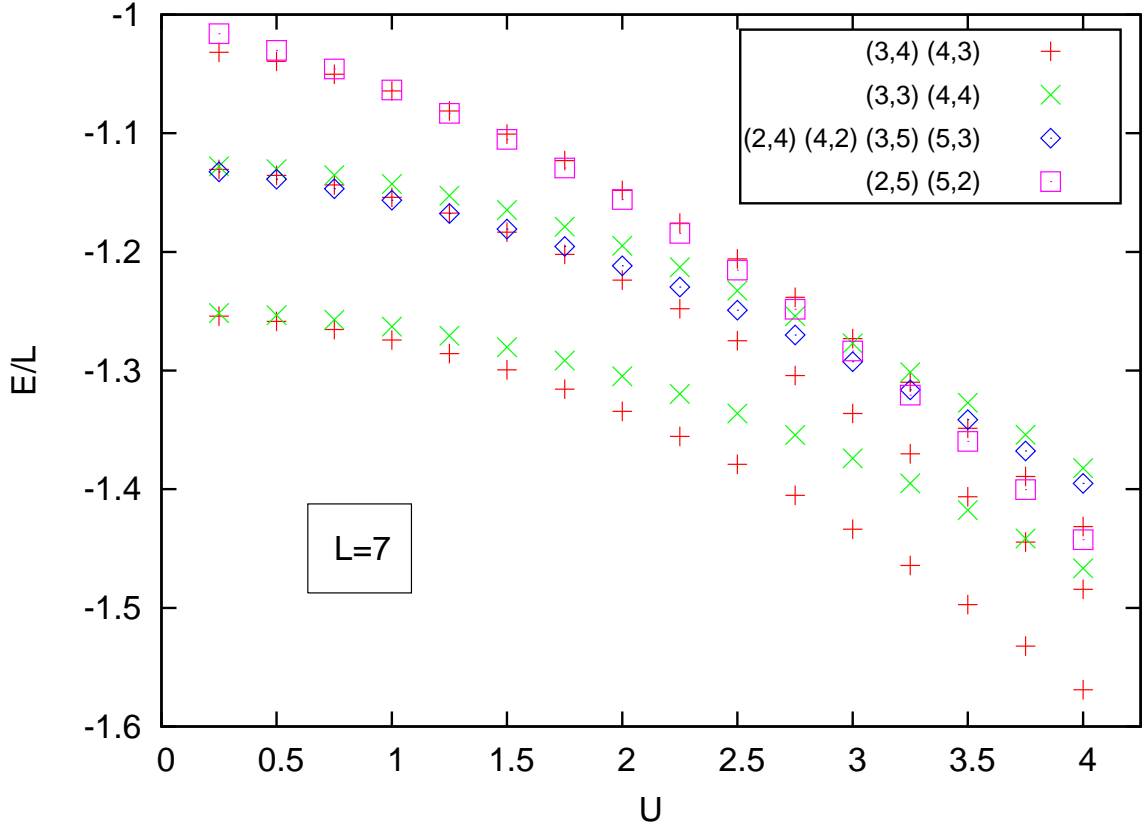}
\end{center}
\caption{The low-lying energies of Hamiltonian (\ref{HAMC}) for $L=7$ as 
function of the interaction $U$. 
The energies are labeled by the quantum numbers  
$(N_{\uparrow},N_{\downarrow})$ of the Bethe 
equations (\ref{Bethe1},\ref{Bethe2}).}
\label{figL7}
\end{figure}

We have performed numerical diagonalization of the charge
pair Hamiltonian (\ref{HAMC}) for small values of odd 
lattice sites.
In Figure (\ref{figL7}) we present the low-lying
energies in the spectrum of the charge pair model for $L=7$.
Considering this analysis we conclude that the ground state sits
in the sectors $(\frac{L+1}{2},\frac{L-1}{2})$ and
$(\frac{L-1}{2},\frac{L+1}{2})$. We find that these states have
zero momenta and consequently the energy of the ground state 
is double degenerated. For sake of comparison we note that these
same states for the Hubbard
model carry non-zero momenta and the respective
energy is therefore four-fold 
degenerated. 

We find that the low-lying states are well 
described 
by real Bethe roots and as before we can take the 
logarithm of the Bethe equations (\ref{Bethe1},\ref{Bethe2}).
Considering the presence of the phase factors we obtain,
\begin{eqnarray}
\label{BetherealODD}
&& L k_j= 2\pi \left[\tilde{Q}_j^{(1)}-\frac{1}{4}\right]-2\sum_{l=1}^{N_{\downarrow}} \arctan\left[\frac{\sin(k_j)-\mu_l}{U/4}  \right],~~~
j=1,\dots,N_{\uparrow}+N_{\downarrow}, \nonumber \\
&& 2\sum_{l=1}^{N_{\uparrow}+N_{\downarrow}} \arctan\left[\frac{\mu_j-\sin(k_l)}{U/4}  \right]=
2\pi \left[\tilde{Q}_j^{(2)}+\frac{1}{2}\right] +2\sum_{\stackrel{l=1}{l \neq j}}^{N_{\downarrow}} \arctan\left[\frac{\mu_j-\mu_l}{U/2}  \right],~~~
j=1,\dots,N_{\downarrow}, \nonumber \\
\end{eqnarray}
where in Table (\ref{TAB6}) we exhibit the numbers 
$\tilde{Q}_j^{(1,2)}$ of selected low-lying states 
which we shall discuss here.
\begin{table}[ht]
\begin{center}
\begin{tabular}{|c|c|c|c|} \hline
$(N_{\uparrow},N_{\downarrow})$ & $\tilde{Q}_j^{(1)}$ & $\tilde{Q}_j^{(2)}$ & $\mathrm{state}$ \\ \hline 
$(\frac{L+1}{2},\frac{L-1}{2})$ & $\frac{L}{2}-j+1$ & -$\frac{(L-1)}{4}+j-1$ & $\mathrm{ground}~\mathrm{state}$ \\ \hline 
$(\frac{L+1}{2},\frac{L-1}{2})$ & -$\frac{L}{2}+j-1$  & $\frac{L-1}{4}-j+1$ & $\mathrm{first}~\mathrm{excitation}$ \\ \hline 
$(\frac{L-1}{2},\frac{L-1}{2})$ & $\frac{L-2}{2}-j+1$ & -$\frac{(L-3)}{4}+j-1$ & $\mathrm{charge}~\mathrm{excitation}$ \\ \hline 
\end{tabular}
\caption{The Bethe numbers $\tilde{Q}_j^{(1,2)}$ for the ground state and two low-lying excitations.}
\label{TAB6}
\end{center}
\end{table}

A distinguished feature of the charge pair model is that it permits us to 
define a mass gap for odd number of sites in analogy to what has been done 
for $L$ even. In fact, from
Figure (\ref{figL7}) we observe that either a hole or a 
particle excitation has the
same energy with respect to the ground state. This leads us to define the following charge gap
for odd number of sites,
\begin{equation}
\label{gapodd}
\Delta_{od}(L)= E_0\left(\frac{L-1}{2},\frac{L-1}{2},U\right)-E_0\left(\frac{L+1}{2},\frac{L-1}{2},U\right)
\end{equation}

We have computed the gap estimates (\ref{gapodd}) by numerically 
solving the 
Bethe equations (\ref{BetherealODD}) for the respective energies 
up to $L=1025$.
The results are exhibited in Table (\ref{TAB7}) and we see that the 
extrapolated estimators are very close to
the exact values (\ref{gap}). 
\begin{table}[ht]
\begin{center}
\begin{tabular}{|c|c|c|c|} \hline
$L$ & $U=2$ & $U=3$ & $U=4$ \\ \hline 
65    &  0.0908120137   &      0.3180826815   &    0.6455305736  \\ \hline
145   &  0.0874329662   &      0.3166431214   &    0.6442188253  \\ \hline
225   &  0.0869930499   &      0.3162724101   &    0.6438819345  \\ \hline
305   &  0.0868183283   &      0.3161057094   &    0.6437310541  \\ \hline
385   &  0.0886720726   &      0.3160118618   &    0.6436463224  \\ \hline
465   &  0.0886658335   &      0.3159520066   &    0.6435923748  \\ \hline
625   &  0.0865835133   &      0.3158804564   &    0.6435279994  \\ \hline
1025  &  0.0865021020   &      0.3158029708   &    0.6434584567   \\ \hline
$ \mathrm{Extrap.}$ & 0.08635(2) & 0.31567(1) &    0.64336(2)  \\ \hline           
$ \Delta(\infty) $ & 0.0863890951 & 0.3156965889  & 0.6433635110  \\ \hline           
\end{tabular}
\caption{The finite-size sequence $\Delta_{od}(L)$ of the charge 
gap (\ref{gapodd}) for
$U=2,3,4$ and the respective extrapolated value. The exact value 
is obtained from expression (\ref{gap}). }
\label{TAB7}
\end{center}
\end{table}

Let us now discuss the behavior of the 
finite-size corrections to the ground state energy. 
The computation of 
these corrections 
can be done within 
the root density formalism \cite{DEV,DEV1} since the 
Bethe equations are solved
by real roots. At this point we recall that this approach has already been 
applied 
to the Hubbard model with even number 
of sites \cite{WAY}. By adapting 
the computations of \cite{WAY} to tackle 
the Bethe equations (\ref{BetherealODD}) 
we find that the 
leading behavior of the
finite-size corrections for the ground state is,
\begin{equation}
\label{ESCAX0}
E_0\left(\frac{L+1}{2},\frac{L-1}{2},U\right)-e_{\infty}L= \frac{2 \pi \xi}{L}\Big(\frac{1}{8}-\frac{1}{12} +\mathcal{O}(1/\ln[LI_0(2\pi/U)]\Big)
\end{equation}
and from the predictions of conformal 
field theory\footnote{We recall that the energy
spectrum of a conformally invariant theory behaves for large $L$ as 
$E_j(L)-e_{\infty}L = \frac{2 \pi \xi}{L}(X_j-\frac{c}{12})$ 
where $c$ is
the central charge and $X_{j}$ 
the respective anomalous dimension \cite{CAR}.}
we conclude that this 
state
has conformal dimension  $X_0=\frac{1}{8}$. 

To support the above result for the scaling dimension we 
compute the 
following finite size two-step estimators for large sizes,
\begin{equation}
\label{DIMX}
X_j(L)=\frac{L}{2\pi \xi} \left[E_j\left(\frac{L+1}{2},\frac{L-1}{2},U\right)-e_{\infty}L\right] +\frac{1}{12}
+\frac{A_j(L)}{\ln[LI_0(2\pi/U)]}
\end{equation}

The strong logarithmic correction in the finite-size 
estimators is considered as follows. 
For each two consecutive values 
of lattice sites we eliminate 
the logarithmic amplitude $A_0(L)$ and calculate the 
respective scaling dimension $X_0(L)$.
In Table (\ref{TAB8}) we present
the results for $X_0(L)$ together with the extrapolated 
value for large $L$. 
We observe that the data
approach the value predicted by 
the root density method $X_0=\frac{1}{8}$ with reasonable precision.
\begin{table}[ht]
\begin{center}
\begin{tabular}{|c|c|c|c|} \hline
$L$ & $U=2$ & $U=3$ & $U=4$ \\ \hline 
65   &   0.1168016878    &       0.1225118676    &     0.12226995302 \\ \hline 
145  &   0.1232924753    &       0.1230064568    &     0.12283310554  \\ \hline
225  &   0.1234848066    &       0.1232046769    &     0.12305775250  \\ \hline
305  &   0.1235761221    &       0.1233224381    &     0.12319059394  \\ \hline
385  &   0.1236397286    &       0.1234039338    &     0.12328221168  \\ \hline
465  &   0.1236879391    &       0.1234652274    &     0.12335091394  \\ \hline
625  &   0.1237522347    &       0.1235460661    &     0.12344131650  \\ \hline
1025 &   0.1238268301    &       0.1236377926    &     0.12354349786  \\ \hline
$ \mathrm{Extrap.}$ & 0.12543(1) & 0.12544(1) & 0.12542(2)  \\ \hline           
\end{tabular}
\caption{The ground state $j=0$ finite-size sequence $X_0(L)$ 
of the conformal dimension (\ref{DIMX})  for 
$U=2,3,4$ and the respective extrapolation for large systems.  
The predicted value for the exponent is $X_0=0.125$.}  
\label{TAB8}
\end{center}
\end{table}

We observe that in analogy to the Hubbard model with $L$ even the scaling 
dimension lacks of 
dependence on the coupling $U$ \cite{WAY}. This fact 
suggests that further insights about the finite-size corrections may
be easily obtained by exploring the
strong coupling limit of the 
Bethe equations (\ref{Bethe1},\ref{Bethe2}). In what follows we will pursue 
this analysis 
for the sector 
with total number of fermions $N_{\uparrow}+N_{\downarrow}=L$ and spin
$N_{\uparrow}-N_{\downarrow}=2n$ where $n$ takes values on half-integers for
$L$ odd.
Formally, this limit may be performed 
by scaling
the spin rapidities as $\mu_j=\frac{U}{2}\lambda_j$ and afterwards 
taking the limit $U \rightarrow \infty$. 
For real momenta $\sin(k_j)$ 
is always bounded and through lowest order in $1/U$ the two-level 
Bethe equations
for the momenta and spin variables decouple. The first Bethe equation (\ref{Bethe1}) 
turn into
a momenta condition for free-fermions while the second level one (\ref{Bethe2}) 
becomes equivalent to that
of the isotropic spin-$\frac{1}{2}$ model with 
twisted boundary condition. More precisely, the equation for the renormalized 
spin rapidities becomes,
\begin{equation}
\Big(\frac{\lambda_l +\frac{i}{2}}
{\lambda_l -\frac{i}{2}}\Big)^{L}=
e^{i\pi }\prod_{\stackrel{k=1}{k \neq l}}^{\frac{L}{2}-n} \frac{{\lambda}_{l}-{\lambda}_{k}+i}{{\lambda}_{l}-{\lambda}_{k} 
-i},~~~l=1, \ldots, \frac{L}{2}-n.
\end{equation}

The critical exponents associated to the spin degrees 
of freedom can 
therefore be inferred from previous analytical and numerical works for the 
spin-$\frac{1}{2}$ Heisenberg, see for
instance \cite{WAY1,HAMER,ALC}. Here we have to combine the
frustrated character of the ground state of the Heisenberg chain 
with the presence of boundary twist. Following \cite{ALC}
and performing the adaption to our situation we find that such
conformal dimensions are\footnote{Note that in \cite{ALC} the vorticity 
are integer numbers while in our
situation they take values on half-integers.},
\begin{equation}
\label{DIMXXX}
\tilde{X}(n,m)= \frac{n^2}{2}+ \frac{(m-\frac{1}{2})^2}{2},~~n=\frac{1}{2},\frac{3}{2},\dots;~~m=\pm \frac{1}{2},\pm \frac{3}{2},\dots,
\end{equation}
where the number $m$ indicates the vorticity of the state.

We note that the ground state scaling 
dimension $X_0=\frac{1}{8}$ coincides 
with the
lowest dimension $\tilde{X}(1/2,1/2)$ of the twisted Heisenberg 
chain. Thus it is plausible to believe that the 
conformal dimensions (\ref{DIMXXX}) should be present in the 
finite-size corrections
of the charge pair model for $L$ odd. In order to give further support 
to this conjecture we now consider the first excitation in the sector 
$(\frac{L+1}{2},\frac{L-1}{2})$ of the charge pair model. This state
has momenta being double degenerated and
the respective logarithmic branch numbers are given 
in the third line
of Table (\ref{TAB6}). By applying the root density 
method to this state we obtain,
\begin{equation}
\label{ESCAX1}
E_1\left(\frac{L+1}{2},\frac{L-1}{2},U\right)-e_{\infty}L= \frac{2 \pi \xi}{L}\Big(\frac{5}{8}-\frac{1}{12} +\mathcal{O}(1/\ln[LI_0(2\pi/U)]\Big)
\end{equation}
whose corresponding conformal dimension is $X_1=\frac{5}{8}$. In Table (\ref{TAB9}) we provide numerical 
support for the above
analytical computation. The extrapolated value is in reasonable accordance with the analytical prediction.
\begin{table}[ht]
\begin{center}
\begin{tabular}{|c|c|c|c|} \hline
$L$ & $U=2$ & $U=3$ & $U=4$ \\ \hline 
65   &  0.6260129719   &   0.6335332494   &  0.6345026359  \\ \hline
145  &  0.6291377527   &   0.6301830740   &  0.6307714643  \\ \hline
225  &  0.6284896233   &   0.6293039768   &  0.6297667556  \\ \hline
305  &  0.6281555616   &   0.6288656639   &  0.6292626234  \\ \hline
385  &  0.6279480092   &   0.6285905082   &  0.6289457396  \\ \hline
465  &  0.6278021291   &   0.6283961414   &  0.6287220234  \\ \hline
625  &  0.6276205639   &   0.6281536039   &  0.6284433391  \\ \hline
1025 &  0.6274273472   &   0.6278947826   &  0.6281470300  \\ \hline
$ \mathrm{Extrap.}$ & 0.62535(1) & 0.62542(1) & 0.62545(2)  \\ \hline           
\end{tabular}
\caption{The first-excitation $j=1$ finite-size sequence $X_1(L)$ of the
conformal dimension (\ref{DIMX}) for 
$U=2,3,4$ and the respective extrapolation for large systems.  
The predicted value for the exponent is $X_1=0.625$.}  
\label{TAB9}
\end{center}
\end{table}

Once again we note the dimension $X_1=\frac{5}{8}$ can be obtained 
either from $\tilde{X}(1/2,-1/2)$ or $\tilde{X}(1/2,3/2)$ in agreement 
with the fact
we are dealing with a momenta state. 
We think that the above arguments strongly suggest that the contributions 
of the spin degrees of freedom
to the finite-size corrections of the charge pair model with $L$ odd are
indeed governed by the conformal dimensions (\ref{DIMXXX}).

We conclude with the following comments. We expect that the 
finite-size behavior 
of the Hubbard model 
for $L$ odd will be different from that described 
above for the charge pair model. First we remark
that the gap definition (\ref{gapodd}) does not apply for the 
Hubbard model because its 
eigenvalues do not satisfy 
the spectral property (\ref{rel}). Besides that exact diagonalization of 
the Hubbard Hamiltonian (\ref{HAM})
reveal us that there exists level crossing among the first 
two lowest energies states
for some finite value of $U$. For
$L=5$ we find that the level crossing is among the ground states 
lying in the sectors $(3,2)$ and $(3,3)$. 
For $L=7$ the crossing occurs for states in the 
sectors $(3,4)$ and $(3,3)$ rather than
among the energies in sectors $(3,4)$ and $(4,4)$. Therefore, the nature of such 
crossings seems to depend
on the parity of the number $(L-1)/2$ and the understanding of the 
large $L$ behavior of the low-lying states
requires further investigation. We plan 
to expand on this preliminary analysis and present it elsewhere since 
most of finite-size results
for the Hubbard model appears to be concentrated on even number of sites.

\section{The Covering Vertex Model}

Here we argue that the fermionic Hamiltonian (\ref{HAMC}) can be derived 
in the context
of the commuting transfer matrix approach \cite{BAX}. We start by recalling the 
Boltzmann weights structure
of the symmetric free-fermion eight-vertex model. The
model has four weights $a,b,c,d$ 
and its Lax operator can be represented as,
\begin{equation}
\label{LAX8}
\mathcal{L}_{0j}=\left[
\begin{array}{cc|cc}
a & 0 & 0 & d \\
0 & b & c & 0 \\ \hline
0 & c & b & 0 \\
d & 0 & 0 & a  \\
\end{array}
\right]_j,
\end{equation}
where the indices $0$ and $j$ refer to the horizontal and vertical spaces of states
of the vertex model, respectively. It is assumed the free-fermion condition among 
the weights,
\begin{equation}
\label{free}
a^2+b^2-c^2 -d^2=0. 
\end{equation}

For d=0 Shastry devised a way
to couple two free-fermion six-vertex 
models by a particular diagonal vertex interaction. As a result
was obtained a new integrable 
vertex model of 
statistical mechanics with non-additive 
$R$-matrix\cite{SHA1,SHA2}. In addition, 
Shastry showed
that the transfer matrix of such model commutes with
the Hamiltonian of an equivalent spin chain derived
from that of the Hubbard model by means of the
Jordan-Wigner transformation,
\begin{equation}
\label{JW}
c_{\uparrow}(j) =\prod_{k=1}^{j-1} \sigma_{k}^{z} \sigma_j^{-},~~~
c_{\downarrow}(j) =\prod_{k=1}^{L} \sigma_{k}^{z} 
\prod_{k=1}^{j-1} \tau_{k}^{z} \tau_j^{-}
\end{equation}
where $\{\sigma_j^{\pm},\sigma_j^{z}\}$ and 
$\{\tau_j^{\pm},\tau_j^{z}\}$  are two commuting sets of Pauli 
matrices acting
the $j$-th lattice site.

In what follows we shall point out that Shastry's approach 
also works in the 
subspace of weights
with $b=0$. Before that we recall that Shastry's construction
has been shown to be applicable when we couple certain
special vertex models invariant under 
the $gl(n|m)$ superalgebra \cite{MASS,MAR2,DRU}. We emphasize
such generalizations
lead to Hamiltonian models with higher number 
of states per site than that of the
charge pair model (\ref{HAMC}) introduced here. The fact that
Shastry's method also works using an eight-vertex model with $b=0$
appears to have been overlooked so far.
For $b=0$ free-fermion condition (\ref{free}) 
is a circle in the affine plane and it 
can be parametrized as,
\begin{equation}
a=1,~~c= \cos(\lambda),~~~
d= \sin(\lambda)
\end{equation}
where $\lambda$ is the spectral parameter. Note that at $\lambda=0$  
the Lax
operator (\ref{LAX8}) becomes a two-dimensional permutator. 

The construction of coupled vertex models for $b=0$ is fairly parallel 
to that
devised by Shastry and in what follows we shall summarize
only the main results. The Lax operator of the coupled model has the
standard Shastry's form,
\begin{equation}
\label{LAXcoup}
\mathrm{{\bf L}}_{0j}(\lambda)=\exp\left[\frac{h(\lambda)}{2}(\sigma_0^{z} \tau_0^{z}+\mathrm{I}_{0})\right] \mathrm{I}_j
\left[\mathcal{L}^{(\sigma)}_{0j}(\lambda) \mathcal{L}^{(\tau)}_{0j}(\lambda)\right]
\exp\left[\frac{h(\lambda)}{2}(\sigma_0^{z} \tau_0^{z}+\mathrm{I}_{0})\right] \mathrm{I}_j,
\end{equation}
where $\mathrm{I}$ denotes the four-dimensional 
identity matrix and $h(\lambda)$ 
characterizes the strength of the
coupling.

In our case, however, the operators
$\mathcal{L}_{0j}^{(\sigma)}(\lambda)$ and 
$\mathcal{L}_{0j}^{(\tau)}(\lambda)$ are two copies of 
the free-fermion eight-vertex
model with $b=0$. These Lax operators can be expressed  
in terms 
of Pauli matrices as,
\begin{eqnarray}
\label{BLOCK}
&& {\mathcal L}^{(\sigma)}_{0j}(\lambda) =
\frac{1}{2}\left[\mathrm{I}_0\mathrm{I}_j + 
\sigma^{z}_{0}\sigma^{z}_{j} \right]+
\cos(\lambda)\left[\sigma^{+}_{0}\sigma^{-}_{j} + \sigma^{-}_{0}\sigma^{+}_{j}\right]+
\sin(\lambda)\left[\sigma^{+}_{0}\sigma^{+}_{j} + \sigma^{-}_{0}\sigma^{-}_{j}\right] \nonumber \\
&& {\mathcal L}^{(\tau)}_{0j}(\lambda) =
\frac{1}{2}\left[\mathrm{I}_0\mathrm{I}_j + 
\tau^{z}_{0}\tau^{z}_{j} \right]+
\cos(\lambda)\left[\tau^{+}_{0}\tau^{-}_{j} + \tau^{-}_{0}\tau^{+}_{j}\right]+
\sin(\lambda)\left[\tau^{+}_{0}\tau^{+}_{j} + \tau^{-}_{0}\tau^{-}_{j}\right]
\end{eqnarray}

As usual the transfer matrix of the respective vertex model 
on the square lattice can be 
written as the 
trace of an ordered product of Lax operators (\ref{LAXcoup}) on the horizontal space,
\begin{equation}
\label{TRA}
T(\lambda) =Tr_{0}[\mathrm{{\bf L}}_{01}(\lambda) \mathrm{{\bf L}}_{02}(\lambda) \dots \mathrm{{\bf L}}_{0L}(\lambda)]
\end{equation}
which gives rise to a family of commuting of transfer matrices 
provide the 
coupling $h(\lambda)$ satisfies
the Shastry's spectral constraint,
\begin{equation}
\label{coup}
\sinh\left[2h(\lambda)\right]= \frac{U}{4} \sin(2\lambda)
\end{equation}

At this point we remark that the 
condition (\ref{free}) with $b=0$ 
and the constraint (\ref{coup}) 
can be translated 
into a single algebraic
relation after a suitable definition of the
ring variables. Indeed, 
following \cite{MR2}
it is possible to define new affine variables,
\begin{equation}
x=c\exp\left[h(\lambda)\right],~~~
y=d\exp\left[h(\lambda)\right]
\end{equation}
such that the spectral curve assuring 
the integrability of the model
is the following genus one quartic curve,
\begin{equation}
\label{CURVE}
(x^2+y^2)^2 -Uxy-1=0
\end{equation}

Now the spin Hamiltonian $H_s$ associated with 
this vertex model 
is obtained by 
expanding the logarithm of the
transfer matrix (\ref{TRA}) around the 
regular point $\lambda=0$. Apart from an additive constant 
its expression
is given by,
\begin{equation}
\label{HAMspin}
H_s =
\sum_{j=1}^{L} \left[\sigma^{-}_{j}\sigma^{-}_{j+1} 
+ \sigma^{+}_{j+1}\sigma^{+}_{j} 
+ \tau^{-}_{j}\tau^{-}_{j+1} + \tau^{+}_{j+1}\tau^{+}_{j} \right]+ 
\frac{\mathrm{U}}{4}\sum_{j=1}^{L}\sigma^{z}_{j}\tau^{z}_{j}, 
\end{equation}
with periodic boundary conditions imposed.

With the help of the Jordan-Wigner transformation (\ref{JW}) 
the fermionic Hamiltonian (\ref{HAMC}) can be rewritten in terms of Pauli operators. It turns out
that this transformation is able to reproduce
only bulk part of the coupled spin chain (\ref{HAMspin}),
\begin{eqnarray}
\label{HAMCC}
H_c &=&
\sum_{j=1}^{L-1} \left[\sigma^{-}_{j}\sigma^{-}_{j+1} 
+ \sigma^{+}_{j+1}\sigma^{+}_{j} 
+ \tau^{-}_{j}\tau^{-}_{j+1} + \tau^{+}_{j+1}\tau^{+}_{j} \right]+ 
\frac{\mathrm{U}}{4}\sum_{j=1}^{L}\sigma^{z}_{j}\tau^{z}_{j} \nonumber \\
&&
-(\sigma_{L}^{-}\sigma_{1}^{-}-\sigma_{1}^{+} \sigma_{L}^{+})\prod_{k=1}^{L-1} \sigma_k^{z}
-(\tau_{L}^{-}\tau_{1}^{-}-\tau_{1}^{+} \tau_{L}^{+})\prod_{k=1}^{L-1} \tau_k^{z}
\end{eqnarray}
since the boundary terms are clearly distinct 
from that of the coupled spin model (\ref{HAMspin}).

In order to match the boundary term we can exploit the fact that 
integrability is
still preserved by performing certain suitable twist 
transformations on
the Lax operators \cite{KUN}.
Besides that we
have to consider that the local states of the 
fermionic Hamiltonian (\ref{HAMC}) are
constituted by a graded space with two bosonic and two
fermionic degrees of freedom. We expect that 
the respective Lax operator at the regular point should be
proportional to the graded permutation operator,
\begin{equation}
P^{(g)}= \sum_{j,k=1}^{4} (-1)^{p_jp_k} e_{jk} \otimes e_{kj}
\end{equation}
where $e_{jk}$ denotes a $4 \times 4$ matrix having only one
non-vanishing element with value 1 at row $j$ and column $k$.
We choose the Grassmann 
parities $p_j$ according
to the basis ordering (\ref{basis}) and therefore we set
$p_1=0,p_2=1,p_3=1,p_4=0$.

Combining the procedures mentioned above we find that 
the suitable fermionic
Lax operator is obtained by the following twist transformation,
\begin{equation}
\label{LAXferm}
{\mathrm{\bf L}}^{(g)}_{0j}(\lambda) =
\mathrm{M}{\mathrm{\bf L}}_{0j}(\lambda) \overline{\mathrm{M}}
\end{equation}
where the twists $\mathrm{M}$ and $\overline{\mathrm{M}}$ are 
the following diagonal matrices,
\begin{eqnarray}
\mathrm{M}&=&\mathrm{diag}(1,1,-1,-1|1,1,-1,-1|1,-1,1,-1|1,-1,1,-1) \nonumber \\
\overline{\mathrm{M}}&=&\mathrm{diag}(1,1,1,1|1,-1,1,-1|-1,1,-1,1|-1,-1,-1,-1)
\end{eqnarray}

It turns out that the explicit matrix representation of the 
Lax operator (\ref{LAXferm}) 
in terms of the spectral variables $x$ and $y$ is given by,
\begin{equation}
\label{LAXfermA}
{\scriptsize
{\mathrm{\bf L}}_{0j}^{(g)}(x,y)=\left(
\begin{array}{cccc|cccc|cccc|cccc}
\omega_1& 0& 0& 0& 0& -y& 0& 0& 0& 0& -y& 0& 0& 0& 0& -y^2 \\
0& 0& 0& 0& x& 0& 0& 0& 0& 0& 0& 0& 0& 0& -xy& 0 \\
0& 0& 0& 0& 0& 0& 0& 0& x& 0& 0& 0& 0& xy& 0& 0 \\
0& 0& 0& 0& 0& 0& 0& 0& 0& 0& 0& 0& x^2& 0& 
0& 0 \\ \hline
0& x& 0& 0& 0& 0& 0& 0& 0& 
      0& 0& \omega_2& 0& 0& 0& 0 \\
y& 0& 0& 0& 0& -1& 0& 
      0& 0& 0& \omega_3& 0& 0& 0& 0& -y \\
0& 0& 0& 0& 0& 0& 
      0& 0& 0& \omega_4& 0& 0& 0& 0& 0& 0 \\
     0& 0& 0& 0& 0& 0& 0& 0& \omega_2& 0& 0& 0& 0& 
      x& 0& 0 \\ \hline
0& 0& x& 0& 0& 0& 0& 
      -\omega_2& 0& 0& 0& 0& 0& 0& 0& 0 \\
     0& 0& 0& 0& 0& 0& \omega_4& 0& 0& 0& 0& 
      0& 0& 0& 0& 0 \\
y& 0& 0& 0& 0& \omega_3 & 0& 0& 0& 0& 
      -1& 0& 0& 0& 0& -y \\ 
     0& 0& 0& 0& -\omega_2& 0& 0& 0& 0& 0& 0& 0& 0& 0& 
      x& 0 \\ \hline
0& 0& 0& x^2& 0& 0& 0& 0& 0& 0& 0& 
      0& 0& 0& 0& 0 \\
0& 0& -xy& 0& 0& 0& 0& x& 0& 0& 0& 0& 0& 
      0& 0& 0 \\
0& xy& 0& 0& 0& 0& 0& 0& 0& 0& 0& x& 0& 0& 
      0& 0 \\
-y^2& 0& 0& 0& 0& y& 0& 0& 0& 0& y& 0& 0& 0& 0& \omega_1 \\
\end{array}
\right)_j,
}
\end{equation}
where the weights $w_1, \dots, w_4$ dependence on the spectral variables are,
\begin{equation}
w_1=x^2+y^2,~~~w_2=\frac{xy}{x^2+y^2},~~~w_3=\frac{-y^2}{x^2+y^2},~~~w_4=\frac{-x^2}{x^2+y^2}.
\end{equation}

We now show that this fermionic Lax operator is able to produce 
the two-body part of the
charge pair Hamiltonian (\ref{HAMC}) through its expansion
around the regular
point $x=1$ and $y=0$. It turns out that the first order expansion 
of the spectral variables 
constrained by the
curve (\ref{CURVE}) is given by,
\begin{equation}
x=1+\frac{U}{4} \epsilon+\mathcal{O}(\epsilon^2),~~~y=\epsilon+\mathcal{O}(\epsilon^2)
\end{equation}
where $\epsilon$ is the expansion parameter. 

Now considering the expansion of the Lax operator (\ref{LAXfermA}) we obtain,
\begin{equation}
{\mathrm{\bf L}}_{jj+1}^{(g)}(x,y) = P^{(g)}\left(1+\epsilon H_{j,j+1}\right)
\end{equation}
where the operator $H_{j,j+1}$ is given by
\begin{eqnarray}
H_{j,j+1}&=&
c_{\sigma}(j)c_{\sigma}(j+1) + c^{\dagger}_{\sigma}(j+1)c^{\dagger}_{\sigma}(j) 
+\frac{U}{2} (n_{\uparrow}(j)-\frac{1}{2})(n_{\downarrow}(j)-\frac{1}{2}) \nonumber \\
&+&\frac{U}{2}(n_{\uparrow}(j+1)-\frac{1}{2})(n_{\downarrow}(j+1)-\frac{1}{2}) +\frac{U}{4} I_j \otimes I_{j+1}
\end{eqnarray}
which coincides with the two-body term of Hamiltonian (\ref{HAMC}) 
apart from a trivial additive factor.

We close this section mentioning that both Lax operators (\ref{LAXcoup},\ref{LAXfermA})  
fulfill the Yang-Baxter relation. This factorization condition together 
with corresponding $R$-matrices 
has been summarized in Appendix A.

\section{Concluding Remarks}

In this paper we have introduced a variant of the Hubbard model 
whose next-neighbor term plays the role of 
a triplet charge pair potential. For arbitrary lattice sizes
the model has two conserved charges which can be added to the 
Hamiltonian
without affecting its integrability. Besides that gauge fluxes can
be attached to both the pair potential and the conserved charges and
an extended charge pair Hamiltonian can be written,
\begin{eqnarray}
\label{HAMGERAL}
{\bf H}_c &=& \sum_{j=1}^{L} \sum_{\alpha=\uparrow,\downarrow} \left[
e^{i\theta_{\alpha}}c_{\alpha}(j)c_{\alpha}(j+1) + e^{-i\theta_{\alpha}}c^{\dagger}_{\alpha}(j+1)c^{\dagger}_{\alpha}(j) \right]
+U \sum_{j=1}^{L}(n_{\uparrow}(j)-\frac{1}{2})(n_{\downarrow}(j)-\frac{1}{2}) \nonumber \\ 
&+&ih_1\sum_{j=1}^{L} \left[
e^{i\left(\frac{\theta_{\uparrow}-\theta_{\downarrow}}{2}\right)}c_{\downarrow}^{\dagger}(j)c_{\uparrow}(j) - e^{-i\left(\frac{\theta_{\uparrow}-\theta_{\downarrow}}{2}\right)}c^{\dagger}_{\uparrow}(j)c_{\downarrow}(j) \right] \nonumber \\
&+&h_2\sum_{j=1}^{L} \left[
e^{-i\left(\frac{\theta_{\uparrow}+\theta_{\downarrow}}{2}\right)}c_{\uparrow}^{\dagger}(j)c^{\dagger}_{\downarrow}(j) + e^{i\left(\frac{\theta_{\uparrow}+\theta_{\downarrow}}{2}\right)}c_{\downarrow}(j)c_{\uparrow}(j) \right]
\end{eqnarray}
where $\theta_{\uparrow},\theta_{\downarrow}$  are
flux phases 
and $h_1,h_2$ are the chemical
potentials associated to the conserved charges.

The fluxes can be removed from the 
Hamiltonian (\ref{HAMGERAL}) by means 
of the canonical transformation 
$c_{\alpha}(j) \rightarrow 
e^{-i\frac{\theta_{\alpha}}{2}}c_{\alpha}(j)$ and 
$c^{\dagger}_{\alpha}(j) \rightarrow 
e^{i\frac{\theta_{\alpha}}{2}}c^{\dagger}_{\alpha}(j)$. The Bethe ansatz 
solution  for
extended Hamiltonian (\ref{HAMGERAL}) follows that 
given in section 3 and the
respective Bethe equations are given by the same
relations (\ref{Bethe1},\ref{Bethe2}). The basic change
is in the expression for the eigenenergies which now is, 
\begin{equation}
\label{EINA}
E(N_{\uparrow},N_{\downarrow},U) = -2\sum_{j=1}^{N_{\uparrow}+N_{\downarrow}} \cos(k_j)+\frac{U}{2}(\frac{L}{2}-N_{\uparrow}-N_{\downarrow})
+h_1(N_{\uparrow}-N_{\downarrow}) +h_2(N_{\uparrow}+N_{\downarrow}-L)
\end{equation}

We have argued that the exact integrability 
of the charge 
pair model (\ref{HAMC}) 
can be established by using
a construction devised by Shastry for the 
Hubbard model \cite{SHA1,SHA2}. This procedure gives rise to an 
equivalent spin chain (\ref{HAMspin}) which can be seen as 
two coupled special $XY$ models. We now show that such spin chain
can be mapped into  
two coupled $XX$ models where the  boundary conditions depend  on 
if we have an even
or odd number of sites.
To this end we define the
following transformation acting on the even sites of the 
lattice,
\begin{equation}
\label{spinTRA}
\sigma_{j}^{\pm} \rightarrow \sigma_j^{\mp},~~\sigma_j^{z} \rightarrow -\sigma_j^{z},~~
\tau_{j}^{\pm} \rightarrow \tau_j^{\mp},~~\tau_j^{z} \rightarrow -\tau_j^{z},~~\mathrm{for}~~j=2,4,6,\dots
\end{equation}

For $L$ even the form of the transformed Hamiltonian (\ref{HAMspin}) is,
\begin{eqnarray}
\label{HAMspinTRAeven}
\tilde{H}_s &=&
\sum_{j=1}^{L-1} \left[\sigma^{-}_{j}\sigma^{+}_{j+1} 
+ \sigma^{+}_{j+1}\sigma^{-}_{j} 
+ \tau^{-}_{j}\tau^{+}_{j+1} + \tau^{+}_{j+1}\tau^{-}_{j} \right] + 
\frac{\mathrm{U}}{4}\sum_{j=1}^{L}\sigma^{z}_{j}\tau^{z}_{j} \nonumber \\
&+& \sigma^{-}_{L}\sigma^{+}_{1} 
+ \sigma^{+}_{1}\sigma^{-}_{L} 
+ \tau^{-}_{L}\tau^{+}_{1} + \tau^{+}_{1}\tau^{-}_{L},~~L=2,4,6,\dots,
\end{eqnarray}
which is exactly the same spin chain associated 
to integrability 
of the Hubbard model \cite{SHA1,SHA2}. The corresponding Bethe 
equations 
have been discussed before \cite{DEG,MAR3} and for
sake of completeness we also present them here,
\begin{eqnarray}
&& e^{ik_jL} = -(-1)^{N_{\uparrow}}
\prod_{l=1}^{N_{\downarrow}}  \frac{\sin(k_j) - {\mu}_{l} + \frac{iU}{4}}{\sin(k_j) - {\mu}_{l} -\frac{iU}{4}},~~~j=1,2,\dots,N_{\uparrow}+N_{\downarrow},
\nonumber \\
&& \prod_{j=1}^{N_{\uparrow}+N_{\downarrow}}  \frac{\sin(k_j) - {\mu}_{l} + \frac{iU}{4}}{\sin(k_j) - {\mu}_{l} - \frac{iU}{4}}  =
(-1)^{N_{\downarrow}+N_{\uparrow}}\prod_{\stackrel{k=1}{k \neq j}}^{N_{\downarrow}} \frac{{\mu}_{l}-{\mu}_{k}-\frac{iU}{2}}{{\mu}_{l}-{\mu}_{k} 
+\frac{iU}{2}},~~~l=1, \ldots, N_{\downarrow},
\end{eqnarray}
where now the phase factors depend on the combined 
parities of the 
quantum numbers of the model. The eigenvalues are once again 
determined by
the expression (\ref{EIN}).

On the other hand when $L$ is odd
the transformed Hamiltonian (\ref{HAMspin}) is given by,
\begin{eqnarray}
\label{HAMspinTRAodd}
\tilde{H}_s &=&
\sum_{j=1}^{L-1} \left[\sigma^{-}_{j}\sigma^{+}_{j+1} 
+ \sigma^{+}_{j+1}\sigma^{-}_{j} 
+ \tau^{-}_{j}\tau^{+}_{j+1} + \tau^{+}_{j+1}\tau^{-}_{j} \right] + 
\frac{\mathrm{U}}{4}\sum_{j=1}^{L}\sigma^{z}_{j}\tau^{z}_{j} \nonumber \\
&+& \sigma^{-}_{L}\sigma^{-}_{1} 
+ \sigma^{+}_{1}\sigma^{+}_{L} 
+ \tau^{-}_{L}\tau^{-}_{1} + \tau^{+}_{1}\tau^{+}_{L},~~L=3,5,7,\dots,
\end{eqnarray}

Now we see that the boundary term in (\ref{HAMspinTRAodd}) breaks 
explicitly 
the two $U(1)$ symmetries present in the
bulk part of the Hamiltonian.  Despite of this fact we found out that 
the transformed model (\ref{HAMspinTRAodd}) still preserves the property of having
factorized reference states associated with the exact eigenvalues 
$E=\pm \frac{LU}{4}$. The situation is similar
to what we have found for the charge pair model 
as shown in Table (\ref{TAB1}). 
The structure of such eigenstates for the spin model (\ref{HAMspinTRAodd})
is however a bit different since it
contain alternating phases in the tensor product. The form of these 
reference
states is summarized in Table (\ref{TAB10}) where $e_j^{(l)}$ 
denote the four dimensional orthogonal vectors acting on the $j$-th
site of the lattice,
\begin{equation}
\label{ORTHO}
{\scriptsize
e_j^{(1)}=
\left(
\begin{array}{cccc}
1 \\ 
0 \\
0 \\
0 \\
\end{array}
\right)_j,~~
e_j^{(2)}=
\left(
\begin{array}{cccc}
0 \\ 
1 \\
0 \\
0 \\
\end{array}
\right)_j,~~
e_j^{(3)}=
\left(
\begin{array}{cccc}
0 \\ 
0 \\
1 \\
0 \\
\end{array}
\right)_j,~~
e_j^{(4)}=
\left(
\begin{array}{cccc}
0 \\ 
0 \\
0 \\
1 \\
\end{array}
\right)_j
}
\end{equation}
\begin{table}[ht]
\begin{center}
\begin{tabular}{|c|c|} \hline
$\mathrm{Eigenvalue}$ & $\mathrm{Spin~Chain~Eigenvector}$ \\ \hline 
$\frac{LU}{4}$ & 
$ \displaystyle \prod_{j=1}^{L} \left(e_j^{(1)} \pm \exp[\frac{i\pi(2j-1)}{2}]e_j^{(4)}\right) $ \\ \hline  
$-\frac{LU}{4}$ & 
$ \displaystyle \prod_{j=1}^{L} \left(e_j^{(2)} \pm \exp[\frac{i\pi(2j-1)}{2}]e_j^{(3)}\right) $ \\ \hline  
\end{tabular}
\caption{Factorized eigenvectors of the transformed coupled 
spin chain (\ref{HAMspinTRAodd}). 
The explicit form of the vectors $e_j^{(l)}$ is given in (\ref{ORTHO}).}
\label{TAB10}
\end{center}
\end{table}

In addition to that we have been able to built few low-lying states 
on top of the  
reference state given in Table (\ref{TAB10}). Carrying on the Bethe ansatz 
analysis for 
such states we end up 
with the same Bethe equations of the charge pair
model, see equations (\ref{Bethe1},\ref{Bethe2}). This strongly suggests 
that the 
eigenenergies
of the charge pair model (\ref{HAMC}) and the coupled
spin chain (\ref{HAMspin}) are the same
for $L$ odd. We have indeed confirmed this fact by comparing 
the spectrum of these 
models with the help
of exact diagonalization for $L=3,5,7$ sites. 
The eigenfunctions structure of such two models should be related
but a more concrete relationship among them has eluded us so far.

Lastly, one characteristic of the one-dimensional charge 
pair model 
is that the 
thermodynamic
limit properties do not depend on fact
that the lattice 
is bipartite. It was argued that the charge gap can be 
defined for even and odd number of sites both converging
to the same value in the infinite size limit. This
should be contrasted to the case of the Hubbard model
in which the lattice bipartiteness plays
important role to stablish certain exact results for
the repulsive interaction in
any lattice dimension \cite{LIEB}.
It seems interesting to investigate whether or not 
the methods used
to obtain significant information for 
the Hubbard model in all dimensions
can also be adapted to the case of
the charge pair model. In particular, if one can
state concrete information for the charge pair model 
in higher dimension
without the need of a bipartite lattice assumption.

\section*{Acknowledgments}
This work was supported in part by the Brazilian 
Research Council CNPq
through the grants 304758/2017-7 and 401694/2016-0.

\addcontentsline{toc}{section}{Appendix A}
\section*{\bf Appendix A: The Yang-Baxter algebra. }
%\label{APENA}
\setcounter{equation}{0}
\renewcommand{\theequation}{A.\arabic{equation}}

A sufficient condition for exact integrability of vertex model is that 
its Lax operator
satisfies the Yang-Baxter equation for some invertible $R$-matrix \cite{BAX}.  
In the case of the Lax operator (\ref{LAXcoup}) this relation can be state as,
\begin{equation}
\mathrm{R}_{12}(\lambda_1,\lambda_2) \mathrm{{\bf L}}_{13}(\lambda_1) \mathrm{{\bf L}}_{23}(\lambda_2)=
\mathrm{{\bf L}}_{23}(\lambda_2) \mathrm{{\bf L}}_{13}(\lambda_1)
\mathrm{R}_{12}(\lambda_1,\lambda_2),
\end{equation}
where the $R$-matrix has the same structure of that proposed by Shastry for the Hubbard model \cite{SHA2},
\begin{eqnarray}
\mathrm{R}_{12}(\lambda_1,\lambda_2)&=&
\cos(\lambda_1+\lambda_2)\cosh\left[h(\lambda_1)-h(\lambda_2)\right]
\mathcal{L}^{(\sigma)}_{12}(\lambda_1-\lambda_2) \mathcal{L}^{(\tau)}_{12}(\lambda_1-\lambda_2) \nonumber \\
&+&\cos(\lambda_1-\lambda_2) \sinh \left[ h(\lambda_1)-h(\lambda_2) \right]
\mathcal{L}^{(\sigma)}_{12}(\lambda_1+\lambda_2) \mathcal{L}^{(\tau)}_{12}(\lambda_1+\lambda_2) \sigma_1^{z} \tau_1^{z}
\end{eqnarray}
except by the fact that the building block operators 
$\mathcal{L}^{(\sigma)}_{12}(\lambda)$ and  $\mathcal{L}^{(\tau)}_{12}(\lambda)$ 
are given by the special eight-vertex models (\ref{BLOCK}).

The Yang-Baxter algebra for the fermionic Lax operator (\ref{LAXfermA}) has similar form,
\begin{equation}
\label{YBF}
\mathrm{R}^{(g)}_{12}(x_1,y_1,x_2,y_2) \mathrm{{\bf L}}^{(g)}_{13}(x_1,y_1) \mathrm{{\bf L}}^{(g)}_{23}(x_2,y_2)=
\mathrm{{\bf L}}^{(g)}_{23}(x_2,y_2) \mathrm{{\bf L}}^{(g)}_{13}(x_1,y_1)
\mathrm{R}^{(g)}_{12}(x_1,y_1,x_2,y_2),
\end{equation}
but now the tensor products in (\ref{YBF}) have to consider the gradation of the three subspaces.

The explicit form of the $R$-matrix turns out to be,
\begin{equation}
{\mathrm{R }_{12}^{(g)}}(x_1,y_1,x_2,y_2)=
{\scriptsize
\left(
\begin{array}{cccc|cccc|cccc|cccc}
{\bf h}& 0& 0& 0& 0& -{\bf d} & 0& 0& 0& 0& -\bf{d} & 0& 0& 0& 0& {\bf a}-{\bf h} \\
0& 0& 0& 0& 1& 0& 0& 0& 0& 0& 0& 0& 0& 0& -{\bf b}& 0 \\
0& 0& 0& 0& 0& 0& 0& 0& 1& 0& 0& 0& 0& {\bf b}& 0& 0 \\
0& 0& 0& 0& 0& 0& 0& 0& 0& 0& 0& 0& {\bf a} & 0& 
0& 0 \\ \hline
0& 1& 0& 0& 0& 0& 0& 0& 0& 
      0& 0& \overline{{\bf b}}& 0& 0& 0& 0 \\
{\bf d} & 0& 0& 0& 0& {\bf q} & 0& 
      0& 0& 0& {\bf q}-{\bf g} & 0& 0& 0& 0& -{\bf d} \\
0& 0& 0& 0& 0& 0& 
      0& 0& 0& {\bf g} & 0& 0& 0& 0& 0& 0 \\
     0& 0& 0& 0& 0& 0& 0& 0& \overline{\bf b} & 0& 0& 0& 0& 
      1& 0& 0 \\ \hline
0& 0& 1& 0& 0& 0& 0& 
      -\overline{\bf b} & 0& 0& 0& 0& 0& 0& 0& 0 \\
     0& 0& 0& 0& 0& 0& {\bf g} & 0& 0& 0& 0& 
      0& 0& 0& 0& 0 \\
{\bf d} & 0& 0& 0& 0& {\bf q} -{\bf g} & 0& 0& 0& 0& 
      {\bf q} & 0& 0& 0& 0& -{\bf d} \\ 
     0& 0& 0& 0& -\overline{\bf b} & 0& 0& 0& 0& 0& 0& 0& 0& 0& 
      1& 0 \\ \hline
0& 0& 0& {\bf a} & 0& 0& 0& 0& 0& 0& 0& 
      0& 0& 0& 0& 0 \\
0& 0& -{\bf b} & 0& 0& 0& 0& 1& 0& 0& 0& 0& 0& 
      0& 0& 0 \\
0& {\bf b} & 0& 0& 0& 0& 0& 0& 0& 0& 0& 1& 0& 0& 
      0& 0 \\
{\bf a} - {\bf h} & 0& 0& 0& 0& {\bf d} & 0& 0& 0 & 0& {\bf d} & 0& 0& 0& 0& {\bf h}  \\
\end{array}
\right),
}
\end{equation}
where the expressions of the matrix entries are,
\begin{eqnarray}
&& {\bf a}=\frac{y_1y_2}{x_1^2+y_1^2}+\frac{x_1x_2}{x_2^2+y_2^2},~~~
{\bf b}=-\frac{x_1y_2}{x_1^2+y_1^2}+\frac{y_1x_2}{x_2^2+y_2^2},~~~ 
\overline{\bf b}=\frac{y_1x_2}{x_1^2+y_1^2}-\frac{x_1y_2}{x_2^2+y_2^2}, \nonumber \\ \nonumber \\
&& {\bf d}=\frac{x_1y_1-x_2y_2}{x_1^2x_2^2-y_1^2y_2^2},~~~
{\bf g} =-\frac{x_1x_2}{x_1^2+y_1^2}-\frac{y_1y_2}{x_2^2+y_2^2},~~~
{\bf h} =\frac{x_1x_2(x_1^2+y_1^2)-y_1y_2(x_2^2+y_2^2)}{x_1^2x_2^2-y_1^2y_2^2},  \nonumber \\ \nonumber \\
&& {\bf q}=\frac{y_1y_2(x_1^2+y_1^2)-x_1x_2(x_2^2+y_2^2)}{x_1^2x_2^2-y_1^2y_2^2}
\end{eqnarray}
such that
$\{x_1,y_1\}$ and $\{x_2,y_2\}$ denote two arbitrary points on the quartic curve (\ref{CURVE}).

We finally recall that it is possible to rewrite the Yang-Baxter 
relation in an alternative form which is insensitive 
to the grading of the spaces \cite{KUL}. With the help of the graded 
permutation one
can define new operators $\mathrm{\check {S}} =P^{(g)} S$ and 
the algebraic relation (\ref{YBF}) becomes,
\begin{equation}
\mathrm{\check {R}}^{(g)}_{23}(x_1,y_1,x_2,y_2) \check{\mathrm{{\bf L}}}^{(g)}_{12}(x_1,y_1) \check{\mathrm{{\bf L}}}^{(g)}_{23}(x_2,y_2)=
\check{\mathrm{{\bf L}}}^{(g)}_{12}(x_2,y_2) \check{\mathrm{{\bf L}}}^{(g)}_{23}(x_1,y_1)
\check{\mathrm{R}}^{(g)}_{12}(x_1,y_1,x_2,y_2).
\end{equation}

\end{document}